# Unveiling the Stable Nature of the Solid Electrolyte Interphase between Lithium Metal and LiPON via Cryogenic Electron Microscopy

**Authors:** Diyi Cheng,[1] Thomas A. Wynn,[1] Xuefeng Wang,[2,*] Shen Wang,[2] Minghao Zhang,[2] Ryosuke Shimizu,[2] Shuang Bai,[2] Han Nguyen,[2] Chengcheng Fang,[1] Min-cheol Kim,[2] Weikang Li,[2] Bingyu Lu,[2] Suk Jun Kim,[3] and Ying Shirley Meng[1,2,4,*]

**Affiliations:**
[1]Materials Science and Engineering Program, University of California San Diego, La Jolla, CA 92121, USA
[2]Department of NanoEngineering, University of California San Diego, La Jolla, CA 92121, USA
[3]School of Energy, Materials and Chemical Engineering, Korea University of Technology and Education, Cheonan 31253, Republic of Korea
[4]Lead Contact

*Co-corresponding author E-mails:
shmeng@ucsd.edu (Y. S. Meng), wangxf.acs@gmail.com (X. Wang)

## SUMMARY

The solid electrolyte interphase (SEI) is regarded as the most complex but the least understood constituent in secondary batteries using liquid and solid electrolytes. The dearth of such knowledge in all-solid-state battery (ASSB) has hindered a complete understanding of how certain solid-state electrolytes, such as LiPON, manifest exemplary stability against Li metal. By employing cryogenic electron microscopy (cryo-EM), the interphase between Li metal and LiPON is successfully preserved and probed, revealing a multilayer mosaic SEI structure with concentration gradients of nitrogen and phosphorous, materializing as crystallites within an amorphous matrix. This unique SEI nanostructure is less than 80 nm and is stable and free of any organic lithium containing species or lithium fluoride components, in contrast to SEIs often found in state-of-the-art organic liquid electrolytes. Our findings reveal insights on the nanostructures and chemistry of such SEIs as a key component in lithium metal batteries to stabilize Li metal anode.

# INTRODUCTION

The past four decades have witnessed intensive research efforts on the chemistry, structure, and morphology of the solid electrolyte interphase (SEI) in Li-metal and Li-ion batteries (LIBs) using liquid or polymer electrolytes, since the SEI is considered to predominantly influence the performance, safety and cycle life of batteries.[1–5] Pioneering work by Peled et al.[6] and Aurbach et al.[7] has independently proposed two widely accepted SEI models – a mosaic SEI and a multilayer SEI – to explain the structural and chemical evolution mechanism during the SEI formation. Regardless of the structural difference in the models, consensus is that most SEIs in organic liquid or polymer electrolytes are comprised of both inorganic species that are thermodynamically stable against Li metal and organic species that are partially reduced by Li metal.[6,7] A recent study using tip-enhanced Raman spectroscopy investigated the nanoscale distribution of the organic species in the SEI formed on amorphous silicon.[8] Although the studies on SEI chemistry and morphology formed by using various electrolyte compositions and electrode materials have been well documented in literature, existing models still require further efforts to be truly validated in terms of the distribution of nanostructures within the SEI layer. The dearth of SEI studies for solid-state electrolytes (SSE) also leaves the SEI formation mechanism at the Li/SSE interphase elusive.

Compared with their liquid-electrolyte analogues, SSEs have drawn increased attention as they promote battery safety[9], exhibit a wide operational temperature window, and improve energy density by enabling Li metal as anode materials for next-generation lithium-ion batteries.[10,11] Despite suitable mechanical properties to prevent Li dendrite penetration,[12] relatively wide electrochemical stability windows,[13] comparable ionic conductivities,[14] and intrinsic safety, most SSEs are found to be thermodynamically unstable against Li metal, where SSE decomposition produces a complex interphase, analogous to the SEI formed in liquid electrolyte systems. Conventional SSEs, including $Li_7La_3Zr_2O_{12}$, $Li_{1+x}Al_xGe_{2-x}(PO_4)_3$, $Li_{10}GeP_2S_{12}$, $Li_7P_3S_{11}$, $Li_{0.5}La_{0.5}TiO_3$ and amorphous lithium phosphorus oxynitride (LiPON) are predicted by DFT thermodynamic calculations to form SEIs upon contact with Li metal due its high reduction potential,[15,16] which have been validated by experimental findings in many cases.[17–23] The nature of these decomposed phases govern the properties of the interface; an ideal passivation layer should consists of ionically conductive but electronically insulating components to prevent the SSE from being further reduced.

As one of the most successful SSEs, LiPON has enabled an all-solid-state thin-film battery with a Li metal anode and a high-voltage $LiNi_{0.5}Mn_{1.5}O_4$ (LNMO) cathode to achieve a capacity retention of 90% over 10,000 cycles with a Coulombic efficiency over 99.98%,[24] indicating the presence of extremely stable interphases between LiPON and electrode materials. The superior electrochemical performance of LiPON against Li metal has attracted numerous research efforts aiming to understand the underlying nature of stable Li/LiPON interphase. Computational efforts calculated the stability window of LiPON against Li metal to be from 0.68 V to 2.63 V, predicting decomposition products in this SEI as $Li_3P$, $Li_2O$ and $Li_3N$.[25] Experimental efforts to identify this stable interphase of LiPON against Li metal, however, have been impeded by the limited characterization techniques available due to the low interaction volume of lithium, the amorphous nature of LiPON, and the extreme susceptibility of both lithium metal and LiPON to ambient air and probe damage.[26,27] Among the limited characterization methodologies, *in situ* X-ray photoelectron spectroscopy (XPS) conducted on LiPON thin films exposed to evaporated Li illustrated chemical change following Li deposition and identified the decomposition products at the Li/LiPON interphase to be $Li_3PO_4$, $Li_3P$, $Li_3N$ and $Li_2O$.[21] Nevertheless, the structure and spatial distribution at nano-scale of these decomposition products and their influence on interfacial stability remain unclear due to the nature of the buried interphase.

Originating from the structural biology field, cryogenic focused ion beam (cryo-FIB) and cryogenic electron microscopy (cryo-EM) have recently been introduced to battery research, and have proven the ability to preserve and probe Li metal for quantitative structural and chemical analysis.[27–29] Li *et al.* observed different nanostructures in SEIs formed in standard carbonate-based electrolyte and fluorinated-carbonate-based electrolyte respectively by using cryo-EM. They hypothesize that the enhanced electrochemical performance using fluorinated electrolyte is attributed to the formation of a multilayer SEI structure, in contrast to the mosaic SEI structure formed with standard carbonate-based electrolyte, which stressed the competing impact of SEI nanostructure versus SEI chemistry for stabilizing Li metal anodes.[30] Further, Cao *et al.* observed a monolithic amorphous SEI in electrolyte that contains highly-fluorinated solvents. The homogeneous and amorphous features of this SEI was proposed to be the key for the largely improved Coulombic efficiency and dense Li plating.[31] These findings highlighted the importance in investigating the SEI nanostructure formed in liquid electrolytes, and also prompted cryo-EM-based examination of the SEI in ASSBs, which can provide missing yet critical insights on how to build a stable interphase between SSE and Li metal.

Given the susceptibility of LiPON and Li under electron beam exposure,[27,32] herein we combined cryo-FIB and cryo-EM to preserve the Li/LiPON interphase and characterize its chemistry and structure. We observed concentration gradients of nitrogen and phosphorous into Li metal, and a <80 nm thick interphase consisting of a distribution of crystalline decomposition products embedded within an amorphous matrix. The observed structural and chemical evolution across the interphase identifies the SEI components to be $Li_2O$, $Li_3N$ and $Li_3PO_4$, with a unique multilayer-mosaic distribution, confirmed by XPS depth profiling. The observed distinct SEI components ($Li_3N$ and $Li_3PO_4$) are compared with the SEI in liquid systems and the multilayer-mosaic distribution sheds light on their positive effect on stable Li metal cycling. According to these findings, we propose a formation mechanism of the interphases and discuss how this type of SEI facilitates stabling cycling against Li metal.

## RESULTS

**Electrochemistry and a methodology for interface sample preparation**

The presence of a stable interface between Li and LiPON was first demonstrated by a thin film battery consisting of a high-voltage spinel LNMO cathode, a LiPON solid electrolyte and a Li metal anode. **Figure 1A** shows the representative voltage curve of LNMO cathode. The discharge capacity did not experience obvious degradation after 535 cycles at a charge/discharge rate of 5C. **Figure 1B** displays the Coulombic efficiency (CE) change as cycling proceeded, where the CE was stabilized beyond 99.70% after 100 cycles and 99.85% after 400 cycles, which is much better than liquid cell equivalents (**Figure S1**). The excellent electrochemical cycling of this full cell not only confirmed the superior cyclability of LiPON against Li metal anode, but also rendered us eligible to further investigate the nature of this stable interfaces within this battery.

To access a buried interface and elucidate its significance for long-term stable cycling, a combination of FIB and TEM was used, as proven effective to explore many interfacial phenomena.[33–36] Considering the high reactivity and beam sensitivity of both Li metal and LiPON, cryogenic protection is necessary to minimize the potential damage and contamination during sample preparation, transfer, and imaging. [9,27,37,38] Here, we have applied a method of transferring a lamella from the bulk sample to the FIB grid in order to avoid the contamination from organometallic Pt or amorphous ice, both of which have been used as bonding materials in FIB.[37,39] Conventionally, organometallic Pt is deposited to attach the lamella to the nano manipulator, which is then transferred to a FIB grid.[37] This process is generally performed at room temperature, but is impeded under cryogenic temperature due to the condensation of organometallic Pt vapor. Alternatively, Zachman *et al*. used water vapor as a connection material, which condenses to amorphous ice in cryo-FIB.[39] Note that both Pt and water can react with Li causing potential damage and artifacts to the Li/LiPON interphase. Therefore, we applied a re-deposition mounting methodology in the cryo-FIB, where the etched Li metal material was trapped and redeposited at the gap between lamella and manipulator or FIB grid as the connection material. Details of this methodology can be found in the Supporting Information and **Figure S2**. In this way, no extra materials were introduced while conducting the cryo-lift-out for preserving and preparing well-defined Li/LiPON interphases for study in this work.

As shown in **Figure 1C**, the Li/LiPON interphase lamella was extracted from a sample that consists of 1.5-μm lithium metal deposited on a LiPON thin film and thinned to less than 120 nm for TEM observation. The TEM sample was transferred into the TEM column with minimum air exposure using a glovebox. The sample protection methods for each transfer process are listed in **Table S2.** Prior to the observation of Li/LiPON interphase, we first examined the beam stability of LiPON in cryo-EM since FIB-prepared LiPON has shown electron beam susceptibility at room temperature.[32] **Figure S3** demonstrates that continuous high-resolution imaging in cryo-STEM did not cause obvious damage or morphology change of LiPON, showing the capability of cryo-FIB and cryo-EM to preserve the structure of otherwise beam intolerant solids.[35] Besides the beam stability, the amorphous phase of LiPON at cryogenic temperature was confirmed by cryo-XRD, as shown in **Figure S4**, to exclude the effect of potentially phase change of LiPON during the following cryo-(S)TEM observations.

**Concentration gradient across the Li/LiPON interface**

**Figure 1D** shows the cryo-STEM dark field (DF) image of the Li/LiPON interface where the Li metal and LiPON regions are approximately distinguished by the contrast difference. Regions were further identified by energy dispersive x-ray spectroscopy (EDS) mapping results of the elemental distribution of P (**Figure 1E**) and N (**Figure 1F**). Interestingly, P and N content were both observed in the Li metal region. To quantify the chemical evolution across the interface, an EDS line-scan was carried out at the region indicated by the black dashed line in **Figure 1D**, where the concentration evolution of P and N were captured and plotted in **Figure 1G**. From the Li metal to the LiPON, both concentrations of P and N had a clear increase and reached their maximum in the bulk LiPON region, where elemental P and N were uniformly distributed in the bulk of LiPON. Interestingly, the presence of P and N was not directly correlated to the contrast difference associated with the Li/LiPON interface. Instead, the concentration changes began from Li metal region, with a gradient of P and N increasing across the interphase. Furthermore, N signal was detected from a much deeper region into the Li metal side than P signal, indicating a further diffusion of N species into the Li metal region compared with P. Based on the concentration gradient in **Figure S5**, the width of Li/LiPON interphase region was about 76 nm.

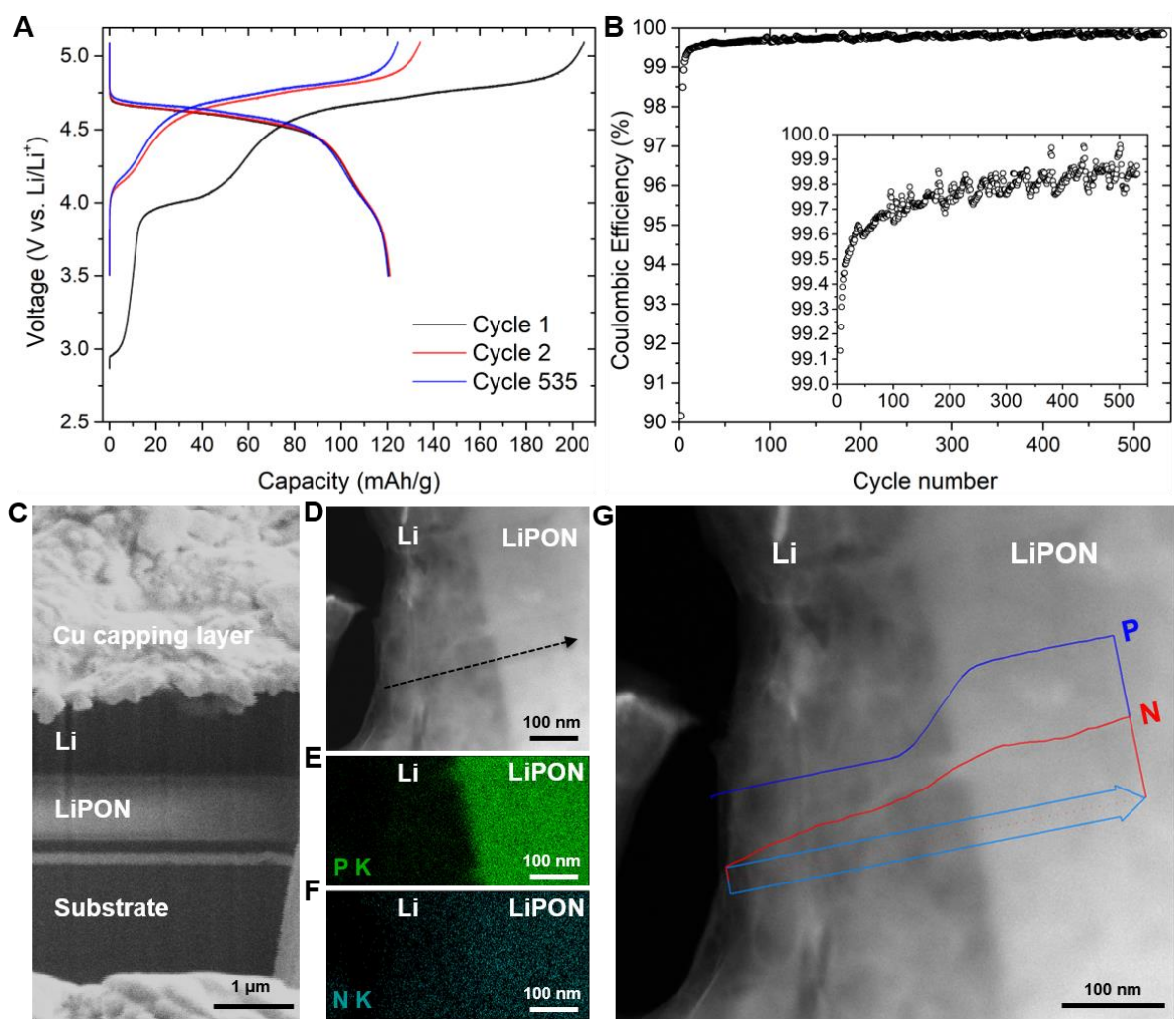

**Figure 1. Electrochemical performance of Li/LiPON/LNMO full cell and cryo-STEM EDS results.** (A) The voltage profiles of the 1$^{st}$, 2$^{nd}$ and 535$^{th}$ cycle. (B) The Coulombic efficiency change with cycle numbers over 500 cycles. (C) Cryo-FIB-SEM cross-sectional image of the Li/LiPON sample. (D) Cryo-STEM DF image of Li/LiPON interface. EDS mapping results of P (E) and N (F) signals in the region shown in (D). (G) EDS line-scan of P and N signals (counts per second) along the black dashed arrow in (D). P and N signals were normalized respectively and plotted along the arrow to show the concentration gradient across the interface. To elaborate the N signal evolution along the interface, EDS spectra at selected spots were shown in Figure S6. In this work, "interface" was used when referring to the physical appearance and position of this Li/LiPON interphase; "interphase" was used when referring to the constituents of the interface and its chemistry or composition.

**High resolution observation of a nanostructured interphase**

To probe the structural evolution associated with the observed concentration gradient, cryo-high-resolution TEM (HRTEM) was performed at the Li/LiPON interface (**Figure 2A**). The inset fast Fourier transform (FFT) pattern in **Figure 2A** first illustrates the coexistence of Li metal, $Li_2O$, $Li_3N$ and $Li_3PO_4$ species distributed in the probing area by matching the lattice spacings of corresponding species with the pattern, hereby identifying this interface as a complex, nanostructured interphase. The compositional evolution from the Li metal region to LiPON region was then investigated stepwise, with FFTs taken from region 1 to region 4 as highlighted by the orange squares in **Figure 2A**, corresponding to **Figure 2B, D, F & H** respectively. In the region near the Li of the Li/LiPON interphase (Region 1), the presence of Li metal and $Li_2O$ were identified based on the FFT spots of (110) plane of Li metal and (111) plane of $Li_2O$ shown in **Figure 2B**. Region 1 represents the beginning of the interphase, with a mixing of Li metal and $Li_2O$, due to the extreme susceptibility of Li metal to oxygen to form $Li_2O$. Moving further inside to Region 2, the FFT (**Figure 2D**) identified the appearance of $Li_2O$, $Li_3N$ and small amount of Li metal, according to the lattice spacings. The (001) FFT spot of $Li_3N$ demonstrated an earlier appearance of N at the interphase, which was likely related to the diffusion of N species within Li metal. Approaching closer to LiPON region (Region 3), no Li metal was observed and $Li_2O$, $Li_3N$ and $Li_3PO_4$ species were identified by FFT shown in **Figure 2F**. All of the species present at Region 3 are considered decomposition products of the LiPON, in part predicted by DFT thermodynamic calculation.[15] **Figure 2H** demonstrates the amorphous structure of LiPON in the bulk region (Region 4).

As for the nanostructures, **Figure 2C** was acquired from Region 1 in **Figure 2A**, where the nanostructures of Li metal and $Li_2O$ were found to be surrounded by amorphous region. The size of these nano crystals was about 3-5 nm. **Figure 2E, G & I** display the nanostructures at the Region 2-4 in **Figure 2A**, respectively. Notably, all the nano crystals were found to be embedded in an amorphous matrix, with a mosaic-like SEI distribution. However, a layered distribution of decomposition products was also indicated as discussed previously from Region 1 to Region 4, which will be further discussed in the following sections. All the nanostructures at the interphase being embedded in an amorphous matrix maintained the fully dense nature of Li/LiPON interphase even after decomposition. From the cryo-TEM and STEM EDS analyses, thus, we observe that (1) the width of Li/LiPON interphase was about 76 nm, (2) the interphase exhibits concentration gradients of P and N, and (3) the interphase consisted of the decomposition products as predicted in the form of nanostructures embedded in a dense amorphous matrix. The presence of $Li_3N$ at the Li/LiPON interphase is analogous to successful liquid-electrolyte SEIs which have enabled stabilized Li metal.[15,40,41] To obtain the statistics of the interphase distribution, the thicknesses of different layers ($Li+Li_2O$, $Li+Li_2O+Li_3N$ and $Li_2O+Li_3N+Li_3PO_4$ layers) within the interphase was extracted from ten different regions, where the depth of each layer was recorded and plotted in **Figure. 2J.** The averaged thicknesses of each layer are summarized in **Figure. 2K**, where the thickness of $Li+Li_2O$, $Li+Li_2O+Li_3N$ and $Li_2O+ Li_3N+ Li_3PO_4$ layers are 21.1 nm, 11.6 nm and 43.7 nm in average, constituting a interphase with an average thickness of 76.4 nm, consistent with the observations from EDS line-scans.

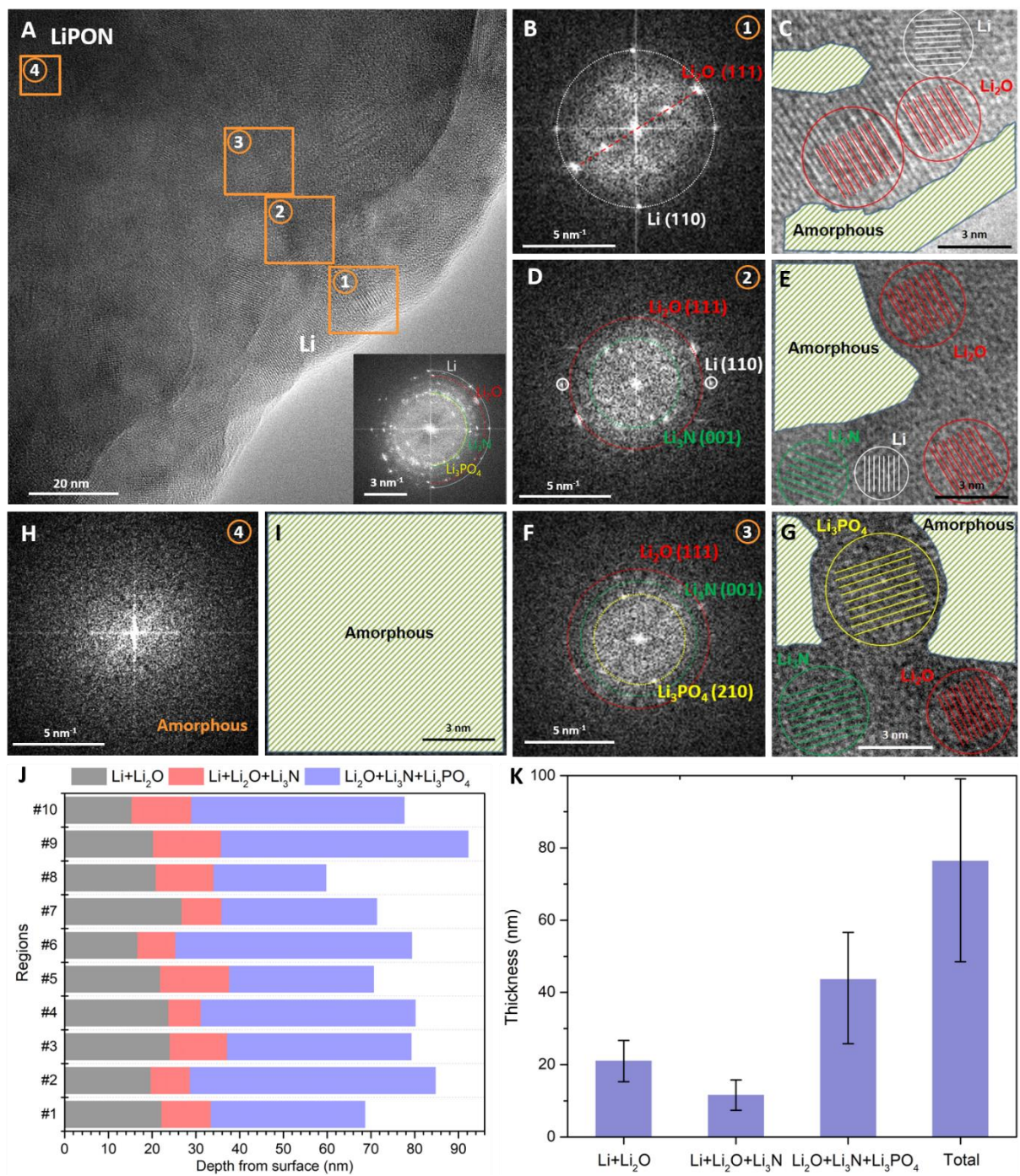

**Figure 2. Nanostructures of Li/LiPON interphase and statistics of cryo-TEM results.** (A) HRTEM image of the interphase where four regions (region 1-4) are highlighted by orange squares to indicate different stages of the multilayered structure across the interphase. Inset image is the FFT result of the whole area in (A). (B, D, F & H) FFT patterns corresponding to region 1-4. (C, E, G & I) Nanostructure schematic corresponding to region 1-4. (J) Depth distribution of different layers within the interphase extracted from 10 different regions. (K) The thicknesses of different layers averaged from the results in (J).

## Cryo-STEM-EELS uncovers local chemical environment

Cryo-STEM-EELS was conducted to obtain further insight of the chemical evolution across the Li/LiPON interphase. **Figure 3A** shows the cryo-STEM DF image of the sample where five spots highlighted within the green arrow were sampled to extract the EELS spectra of Li K-edge, P L-edge and O K-edge along the interphase shown in **Figure 3B**. EELS spectra were acquired every 12 nm with the lowest point located at the LiPON region. As comparison, EELS spectra for corresponding edges of $Li_2O$, $Li_3PO_4$, $Li_3P$ and LiPON species were simulated by FEFF9 and shown in **Figure 3C**. The amorphous LiPON structure (shown in **Figure S7**) was generated by *ab initio* molecular dynamics (AIMD) following the protocol outlined by Lacivita *et al.*[42]

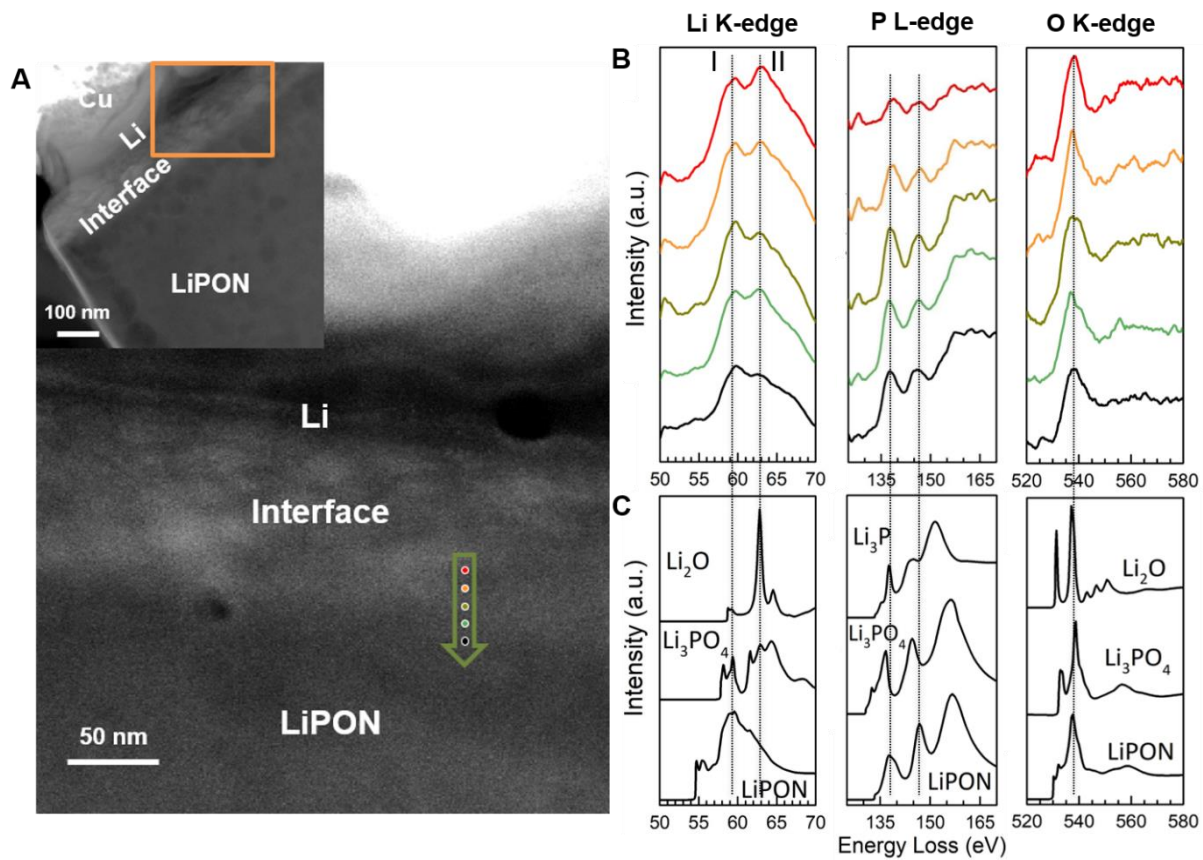

**Figure 3. Cryo-STEM-EELS analysis of Li/LiPON interphase.** (A) Cryo-STEM DF image of Li/LiPON interphase, where five spots highlighted in the green arrow are sampled to extract EELS spectra of Li K-edge, P L-edge and O K-edge shown in (B). The spacing between each sampling point is 12 nm. Inset is a low-magnification STEM DF image of the sample, where the orange rectangle indicates the sampling area shown in the main image. (C) Li K-edge, P L-edge and O K-edge EELS spectra of $Li_2O$, $Li_3P$, $Li_3PO_4$ and LiPON simulated by FEFF9.

The experimentally measured EELS spectra for Li K-edge, P L-edge and O K-edge at LiPON region (black spectra in **Figure 3B**) agreed well with the simulated EELS spectra for corresponding edges of LiPON in **Figure 3C**. The consistency further corroborates the structural model used to generate LiPON EELS, which has been unclear until recently.[26,42,43] The two main peaks as labeled as peak I and peak II in the Li K-edge spectra in **Figure 3B** have brought intriguing insights. According to the simulation, peak I (located at around 59.5 eV) corresponded to the main peak in Li K-edge spectra of LiPON while peak II (located at around 63 eV) corresponded to the main peak of $Li_2O$ (**Figure 3C**). As moving from the interphase towards LiPON region, the intensities ratio of peak I to peak II increased in the experimental spectra. This implied that both $Li_2O$ and LiPON contributed to the experimental Li K-edge spectra (**Figure 3B**), and that the contribution from $Li_2O$ was decreasing as approaching closer to LiPON region. This observation agreed with the cryo-TEM results, where the interphase was identified as nanocrystals distributed within an amorphous matrix that is likely to consist of structural units of LiPON. In terms of P L-edge spectra in **Figure 3B**, the two peaks located around 138 eV and 141 eV originated from the P-O polyhedral structure, which were consistent with peak features in the simulated P L-edge from $Li_3PO_4$ and LiPON in **Figure 3C**, as both have P-O/N polyhedra as the primary structural units. No obvious changes of the edge features were observed except the peak intensities for the P L-edge from interphase to LiPON, indicating the presence of P-O polyhedra at the interphase, emphasizing its structural stability. Similarly for the O K-edge, the experimental spectra did not exhibit notable changes in the edge features through the interphase, indicating the persistence of the local structure in the form of P-O polyhedra. Thus, cryo-STEM-EELS confirmed that the decomposition products were embedded in the amorphous matrix, which was likely to be a mixing of P-O tetrahedrons.

**Chemical evolution confirmed by XPS depth profiling**

Cryo-EM analysis revealed the structure and chemistry of the nanoscale Li/LiPON interphase, though locally. To complement the observation from cryo-EM in a larger scale across the interphase and confirm the structural distribution within the SEI structure, X-ray photoelectron spectroscopy (XPS) depth profiling was conducted on Li/LiPON thin films samples with 100-nm-thick Li metal evaporated on the top of the LiPON. Since the etching rate was non-quantitative, the etching depth was linearly converted from the etching time and thus shown with an arbitrary unit. **Figure 4** illustrates the chemical evolution of O 1s, N 1s, P 2p and Li 1s regions of the Li/LiPON sample with etching through the interphase layer. For comparison, reference XPS spectra of a LiPON thin film sample was shown in **Figure S8**. Before etching started, only O 1s and Li 1s signal were obtained, which can be attributed to the surface $Li_2CO_3$ and interphase $Li_2O$ species. At an etching depth of 54, signal from N 1s appeared. As the N 1s peak became stronger, the spectra could be assigned to $Li_3N$, appearing at a binding energy of 394.4 eV. Note that no P 2p signal was detected at this stage. However, when the etching depth reached 138, the presence of P 2p peak located at 132.8 eV implied the existence of P-containing species, which was mainly attributed to phosphate groups at the interphase. The composition content changes of O 1s, N 1s and P 2p were plotted in **Figure S9**, showing that after the C signal was mostly eliminated at the etching depth of 54, the content of Li and O almost remained the same along the interphase. The concentration gradient of N and P species were also present where N 1s signal appeared first during etching and P 2p signal started to emerge after. The unique sequential distribution of O-, N- and P- containing species identified by XPS depth profiling provided another evidence of the multilayered structure of such interphase, where mosaic structures were present in each layer, according to the cryo-EM findings.

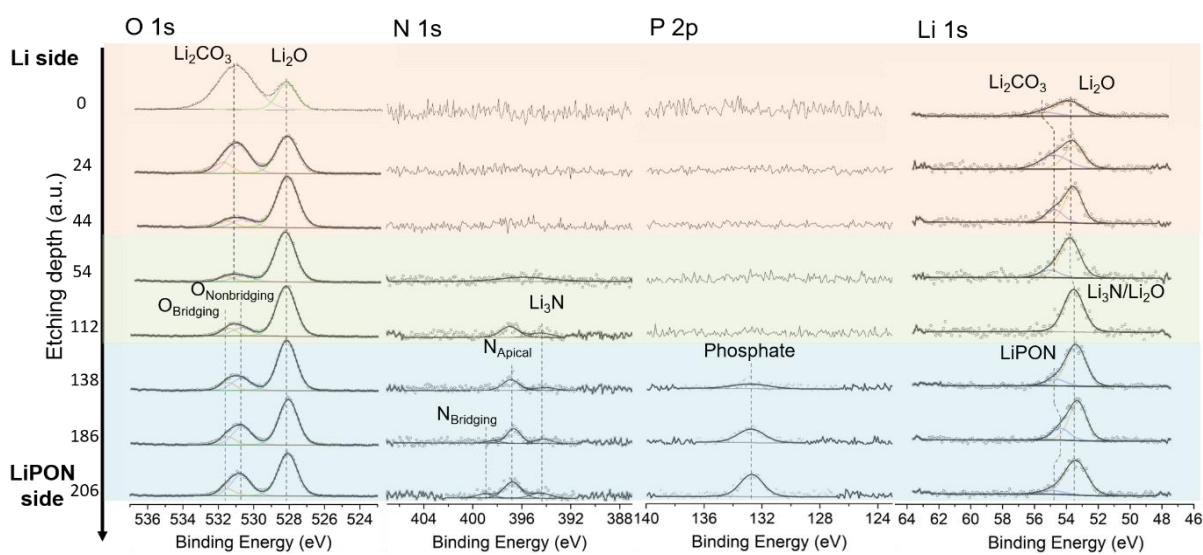

**Figure 4. XPS analysis.** Chemical evolution of O 1s, N 1s, P 2p and Li 1s along Li/LiPON interphase by XPS depth profiling.

## DISCUSSION

**On the formation of a stable interphase**

Through cryo-EM, spatially-resolved characterization of a solid-solid interphase on the order of 100 nm was achieved, highlighting the importance of precise control of temperature and environment when observing buried interfaces. The formation of such a fine interphase requires consideration of potential mechanistic pathways for decomposition, but also stabilization. A primary consideration, recent literature has described RF-sputtered LiPON as a dense, stable glass. Sputtered glassy films are desirable for their uniformity, but also their high density, potentially exhibiting characteristics of a glass annealed on extremely long time-scales. So-called ultrastable glasses exhibit high kinetic stability and are speculated to be one source of the remarkably small interphase.[44]

Despite the potential for kinetic stability, the high reduction potential of Li metal will drive decomposition of a pristine SSE interface, predicted by DFT. By alternating the composition and chemical potential of Li, one can compute a grand potential space where a convex hull can be constructed. Compounds that sit on the convex hull at a given lithium chemical potential are considered stable against Li metal.[16] These DFT results suggested a decomposition reaction between Li and LiPON will result in the formation of $Li_3P$, $Li_2O$ and $Li_3N$ as the equilibrium constituents, which has been complemented by *in situ* XPS findings.[15,21] At high potentials alternate phase equilibria are predicted, forming $P_3N_5$, $Li_4P_2O_7$, and $N_2$ at the oxidation potential;[15] these results are counterintuitive, provided the cyclability of cells including the Li/LiPON interface, and suggest other considerations are lacking, particularly compositional variability.

Converse to calculated phase equilibria, we observe the Li/LiPON interphase to consist of $Li_3PO_4$, $Li_2O$, $Li_3N$ within an amorphous matrix, lacking a clear signature of $Li_3P$. While predicted[16] and observed,[20,45] $Li_3P$ is unlikely to be stable at an interphase at equilibrium. The metastability of Li$_3$P is further corroborated by its absence at the Li-metal/Li$_7$P$_3$S$_{11}$ interphase.[46] These observations highlight the potential difference between metastable, transient states, as likely observed via *in situ* XPS, and equilibrium structures achieved by thick layers of Li. These differences may be brought on by the modified activity of reduced volumes of Li metal. Diffusivity of decomposed ions also provides chemical flexibility in stable phase formation, here, driven by the low formation energy of Li$_3$PO$_4$ relative to Li$_3$P (-2.769 eV and -0.698, respectively).[47]

Further deviation from the predicted phase equilibria exist as a function of spatial distribution of the nanocrystals. This is likely enabled by the surprisingly wide distribution of N and P signatures through the interface, suggesting the concurrent dissociation via the reductive potential of Li and elemental diffusion due to the presence of elemental concentration gradients. Under-coordinated apical N (N$_a$) sites are most susceptible to bond dissociation, exhibiting bond strengths nearly half as low as P-O bonds.[48] Cleavage energy calculations of P-N$_a$ and P-O bonds from isolated phosphate tetrahedra similarly show P-O bonds in isolated PO$_4$ polyhedra to be approximately three times stronger than P-N, with N bridging (N$_b$); this is consistent with previous literature showing that P-N$_b$ bond tends to be the first chemical bond to break when LiPON is reduced by Li metal.[49] After the cleavage of P-N$_b$ bond, the remaining undercoordinated PO$_3$ either give way to further decomposition or contribute to the formation of the amorphous matrix.

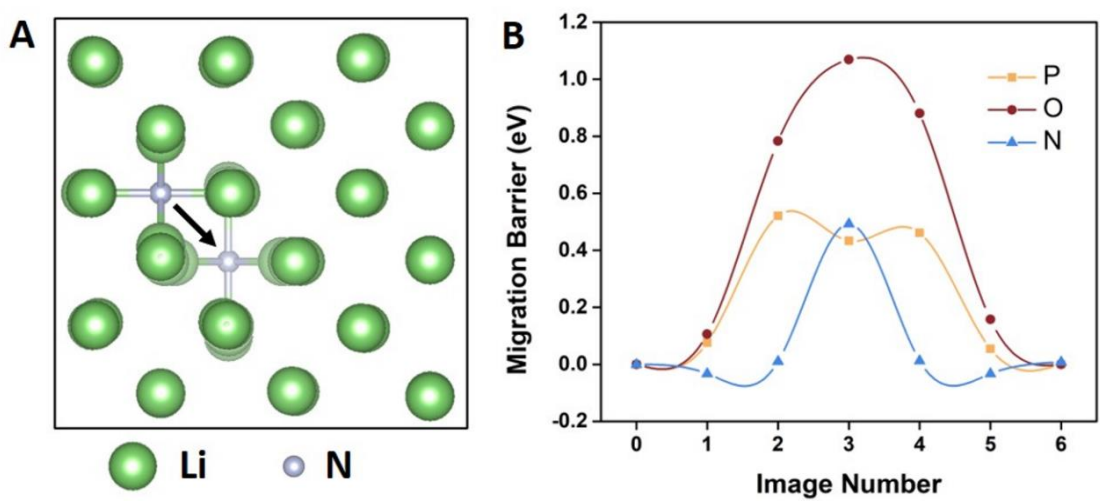

**Figure 5. Interstitial diffusion.** Decomposed LiPON components may (A) diffuse through Li metal via interstitial diffusion. Transition state calculations show (B) diffusion barriers of P, O and N in Li metal to be 0.42 eV, 0.5 eV and 1.08 eV, respectively, indicating the low diffusion barriers of N and P.

The presence of the N and P gradients through the interface (as determined by dark field contrast) indicates that there is significant diffusivity of the decomposed species within the Li metal. To corroborate the potential for diffusion through Li metal, complimentary transition state calculations show a low energy barrier for interstitial diffusion (**Figure 5A**) for both P and N (0.42 eV and 0.5 eV, respectively, as shown in **Figure 5B**), a likely contribution to gradients observed via EDS. While it is known that nitrogen incorporation into Li$_3$PO$_4$ structure enhances the ionic conductivity by two orders of magnitude[43], previous computation efforts using either bulk crystalline LiPON structure [23] or LiPON chains [49] against Li metal, showed that P-N-P bond at the bridging-N site is the most thermodynamically and kinetically unstable in LiPON structures.

It should become apparent that the presence of a complex, stable interphase enabling stability against Li metal in part by the modification of the phase diagram associated with decomposition. At a very early stage, Li metal first reacts with LiPON and diffuses into LiPON region in the form of $Li^+$ ions. With the proceeding of the interphase equilibration, decomposed structural units or ions will remain mobile within Li metal either diffusing through the bulk of Li metal or crystallizing when a critical concentration is achieved. Diffusion gradients observed result in the gradual shielding of the SSE, ultimately reducing the reductive potential of Li acting on the LiPON. As concentrations of dissociated atoms increase at the interface, further structural reconfiguration may occur, where P combines with surrounding undercoordinated Li, O to form a more stable $Li_3PO_4$ instead of $Li_3P$, as the XPS gives primarily the phosphate signal in P 2p region at the interphase. The interfacial decomposition and reconfiguration result in the formation of an 80-nm-thick interphase with N and P gradients between Li metal and LiPON. *In situ* approaches are required for validating the proposed formation mechanism.

In short, the formation of this stable interphase is likely a unique combination of kinetic stability of the glassy electrolyte and the decomposition of highly diffusive species within Li metal that form a variety of nanocrystals within an amorphous matrix. Evolution of this interphase under electrochemical stimuli will be reported in follow up work.

**A distinctive SEI structure found at Li metal/LiPON interphase**

Characterization results obtained from cryo-EM methodology have raised some intriguing insights from the Li/LiPON interphase. Concentration gradients of P and N are present across the interphase. The decomposition products, $Li_2O$, $Li_3N$, $Li_3PO_4$ and an amorphous matrix, were clearly identified at the interphase with a length of about 76 nm and a multilayer-mosaic SEI component distribution. Since $Li_2O$, $Li_3N$ and $Li_3PO_4$ appear as equilibrium phases at the interface of Li metal, such a thin interphase with ionically conductive but electronically insulating components in a gradient configuration is capable of reducing the effective activity of the Li metal anode, shielding the solid electrolyte from further decomposition, as demonstrated in **Figure 6**. Such an eminent passivating effect cannot be realized when the decomposition products from SSEs are mixed electronic and ionic conductors. For instance, $Li_{10}GeP_2S_{12}$ and $Li_{0.5}La_{0.5}TiO_3$ produce electronically conductive Li-Ge alloy and titanates upon being reduced by Li metal that are not able to alleviate the continuous decomposition.[15] In contrast, a similar passivation layer that consists of LiCl, $Li_2S$ and reduced phosphorous species has been identified between Li metal and $Li_6PS_5Cl$ to account for the good cyclability of $Li_6PS_5Cl$ against Li metal anode.[46] However, given the physical properties of different SEI components (**Table S2**), $Li_3N$ and $Li_3PO_4$ are likely to be more suitable for constituting a good SEI than LiCl or $Li_2S$, due to their higher ionic conductivity and lower electronic conductivity.

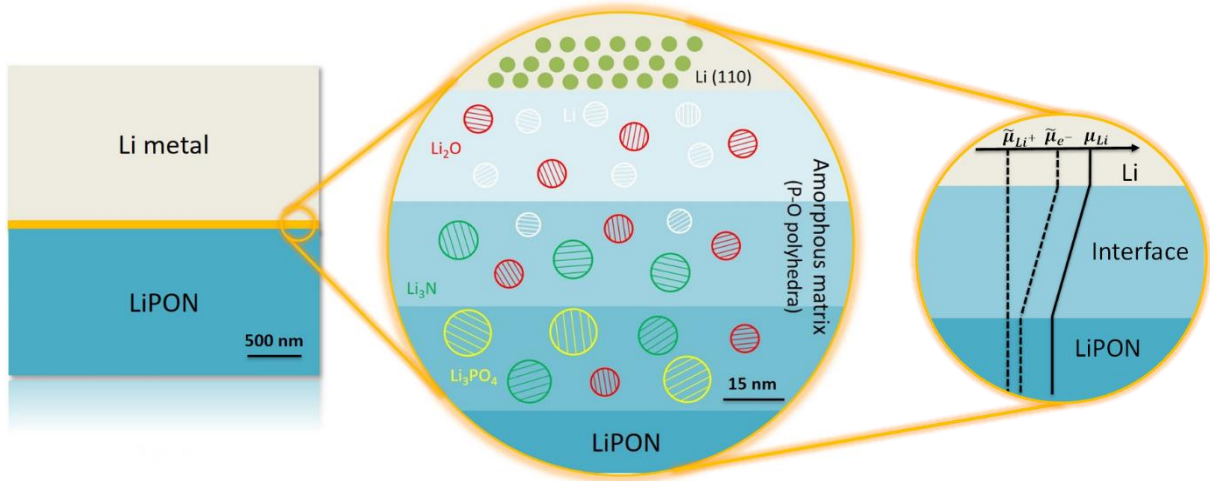

**Figure 6. Li/LiPON multilayered interphase schematic.** $\tilde{\mu}_{Li^+}$, $\tilde{\mu}_{e^-}$ and $\mu_{Li}$ are the electrochemical potential of Li ion, the electrochemical potential of electron and the chemical potential of Li, respectively.

From a perspective of successful SEI composition, major components within SEIs formed in liquid electrolytes consist of $Li_2O$, $Li_2CO_3$, LiF and other alkyl lithium species that are partially reduced by Li metal during SEI formation. The prevalent belief in the passivating effects of such SEIs has driven numerous research efforts to elucidate the passivation mechanism of these species. Nevertheless, the poorly understood alkyl lithium species within the SEIs makes the exact roles of inorganic species including $Li_2O$, $Li_2CO_3$ and LiF unclear. As one of the most popular SEI components that has been extensively studied, LiF, is known for its low electronic conductivity and high thermodynamic stability against Li metal as to explain its passivation effects against Li metal.[15] However, its view as the dominant contribution in Li stabilization has been questioned in a recent work.[50] In the SEI known for its good electrochemical stability, there are only inorganic species present at Li/LiPON interphase, which could also raise concerns regarding truly validating the roles of LiF on constructing a good SEI when all the alkyl lithium species are absent. As has been proposed previously, $Li_3N$ is one of the most promising candidates as an SEI component, due to its thermodynamic stability against Li metal, high ionic conductivity and extremely low electronic conductivity.[40,51] Consequently, the presence of $Li_3N$ at Li/LiPON interphase accounts for the good cyclability of LiPON against Li metal to some extent. From a perspective of SEI structure, the fact that the decomposition products exist as nanostructures and are embedded in a dense amorphous matrix in a mosaic form brings another important perspective of a good interphase for Li metal, where there are no porosity or grain boundaries present that may become nucleation sites for dendrite growth.

## CONCLUSION AND OUTLOOK

In summary, we successfully preserved and characterized the Li/LiPON interphase by developing the cryo-lift-out methodology and combining cryo-FIB and cryo-S/TEM. The observed 76-nm-thick Li/LiPON interphase consisted of SEI components including $Li_2O$, $Li_3N$ and $Li_3PO_4$, which remained fully dense after decomposition. We discovered the concentration gradients of N and P species along the interphase, consistent with the structural evolution identified by cryo-HRTEM. A multilayer-mosaic SEI model was proposed based on these observations. We further proposed the reaction mechanism for Li/LiPON interphase, stressing the diffusion of decomposition product species and structural reconfiguration during equilibration. The comparison with SEIs formed in liquid electrolyte raised questions regarding the roles of alkyl lithium species and LiF in stabilizing Li metal. We caution that electrochemical stability of the Li/LiPON interface, while of utmost importance, explains only some of the high voltage cyclability and LiPON remains one of the few SSEs that can withstand the oxidative potential present in cycling with high voltage cathodes, in stark contrast to liquid electrolyte counterparts and emphasizing the importance of complementary cathode-electrolyte interphase characterization. Nevertheless, the observed structure and proposed mechanistic pathway provide valuable insights for further study on other solid interphases in battery systems by both computational and experimental efforts, raising the importance of kinetics in the modification of phase diagrams, and giving rise to a better understanding of the stability of such interphases. A good interphase needs to fulfill several requirements to obtain exemplary cyclability - formation of a stable passivation layer, uniform coverage, fully dense and thermodynamic stability with Li metal. While an ideal SEI has yet to be demonstrated with liquid electrolytes, LiPON fills these requirements and exemplifies stable Li metal cycling, paving the way towards high-energy long-standing batteries.

## Experimental Procedures

**Resource Availability**

*Lead contact*

Further information and requests for resources and materials should be directed to and will be fulfilled by the Lead Contact, Ying Shirley Meng (shmeng@ucsd.edu).

*Materials Availability*

This study did not generate new unique materials.

*Date and Code Availability*

This study did not generate or analyze (datasets or code).

**Sample preparation**

LiPON thin film was deposited on Pt/Cr/SiO$_2$/Si substrate by radio-frequency (RF) sputtering using a crystalline Li$_3$PO$_4$ target (2″ in diameter, from Plasmaterials, Inc.) in UHP nitrogen atmosphere. Base pressure of the sputtering system was 3×10$^{-6}$ Torr. LiPON deposition used a power of 50W and nitrogen gas pressure of 15 mTorr. The as-deposited LiPON thin film was 1 μm in thickness with a growth rate of ~0.46 Å/min. Ionic conductivity of as-deposited LiPON thin film was measured by electrochemical impedance spectroscopy (EIS) to be 3×10$^{-6}$ S/cm, similar to that in literature.[26] After RF sputtering, LiPON thin film was transferred with environmental isolation from the sputtering chamber to thermal evaporation chamber to minimize air exposure and prepare for lithium metal deposition. Lithium metal thin film was evaporated on LiPON in a high-vacuum chamber with a base pressure lower than 3×10$^{-8}$ Torr. Growth rate and film thickness of the lithium metal were monitored by a quartz crystal microbalance (QCM). The average evaporation growth rate was calibrated to be ~1.53 Å/s. Film thickness was controlled by deposition rate and deposition time. For the full cell fabrication, LNMO cathode was first deposited on Pt-coated (100 nm thick) alumina substrate by pulsed laser deposition (PLD) using a Lambda Physik KrF Excimer laser. Laser fluence and repetition rate were set at ~2J cm$^{-2}$ and 10 Hz. During deposition, substrate temperature was 600°C and oxygen partial pressure was controlled at 0.2 Torr. LNMO film had a thickness of 650 nm with an active area of 4.9 mm$^2$ and an active mass of ~0.013 mg. Active material loading is 0.03 mAh/cm$^2$. LiPON solid electrolyte and Li metal anode were subsequently deposited following the procedures above. The thickness of Li metal anode was 570 nm, which corresponded to 203% excess capacity compared to that of cathode.

**Liquid cell fabrication**

The materials were all purchased from vendors without any further treatment. The electrode was casted on the Al foil by the doctor blade method. The ratio of active material (LNMO, Haldor Topsoe), conductive agent (SPC65, Timical) and binder (PVDF HSV900, Arkema) was 90:5:5, the electrode was dried in the vacuum oven overnight after casting. Active material loading is 0.65 mAh/cm$^2$ (~4.5mg/cm$^2$). The size of the electrode was 12.7mm as the diameter, the coin cell type was CR2032. 50 μL electrolyte (1M LiPF$_6$ in EC:EMC=3:7 wt%), one piece of Celgard 2325 separator and Li metal chip were used. As for the testing protocols, two cycles at C/10 (1C = 147 mA/g) were applied and rest cycles were conducted at C/3.

**Electrochemical measurement**
Thin film full cell was cycled between 3.5 V and 5.1V using a Biologic SP-200 low current potentiostat. A constant current of C/10 was applied at the 1$^{st}$, 2$^{nd}$ and 535$^{th}$ cycle. A constant current of 5C was applied during the rest of cycles.

**Cryogenic focused ion beam/scanning electron microscopy (cryo-FIB/SEM)**
A FEI Scios DualBeam FIB/SEM equipped with cryo-stage was used to observe the surface and cross-section morphology of Li/LiPON sample and prepare for TEM sample. The operating voltage of electron beam was 5 kV. Emission current of electron beam was set to 25 pA to minimize potential damage of electron beam on Li/LiPON sample surface and cross-section. An argon ion beam source was used to mill and thin the sample. The operating voltage of ion beam source was 30 kV. Different emission currents of ion beam were chosen for different purposes, i.e. 10 pA for imaging by ion beam, 0.1 nA for cross-section cleaning/lamella thinning and 3 nA for pattern milling. To preserve the Li/LiPON interphase during TEM sample preparation, cryo-stage was used during pattern milling, cross-section cleaning and lamella thinning processes, where the temperature of cryo-stage was maintained at around -185°C due to heat exchanging with cooled nitrogen gas.

**Cryogenic lift-out methodology**
Conventional cryo-FIB preparation process requires the stage and sample to cool down and remain stable at liquid nitrogen temperature before further milling or thinning, which alone approximately consumes at least 1.5 hour and about 5 liters of liquid nitrogen. Pt deposition was required to connect lamella with the tungsten probe for lamella lift-out and mounting, which could not be performed due to the inability to heat Pt source under cryogenic temperature (around 100K). To avoid repeatedly cooling and warming the stage during Li/LiPON TEM sample preparation, we applied a cryo-lift-out methodology by using redeposition, which has essentially improved the work efficiency and saved research resources. **Figure S2** demonstrates the methodology to complete a cryo-lift-out without Pt deposition at liquid nitrogen temperature, which saves 3-4 hours for each TEM sample preparation.

**Cryogenic (scanning)/transmission electron microscopy (cryo-S/TEM)**
The Li/LiPON interphase lamella for cryo-EM observation was extracted from a separate deposition, which was comprised of Li metal, LiPON and substrates. The Li/LiPON lamella was transferred from the FIB chamber under vacuum using an air-free quick loader (FEI), and stored in an Ar purged glovebox. STEM/EDS line scan results and TEM images were recorded on a JEOL JEM-2800F TEM, equipped with a Gatan Oneview camera operated at 200 kV. A single-tilt liquid nitrogen cooling holder (Gatan 626) was used to cool the samples to approximately -170°C to minimize electron beam damage where the TEM grids were sealed in heat-seal bags and transferred to TEM column using a purging home-made glovebox filled with Ar gas. STEM/EELS results were obtained on a JEOL JEM-ARM300CF TEM at 300 kV. A TEM cryo-holder (Gatan) was used to load the sample where TEM grids were immersed in liquid nitrogen and then mounted onto the holder via a cryo-transfer workstation. The whole

TEM sample preparation and transfer process guaranteed minimum contact of Li metal with air.

**X-ray photoelectron spectroscopy**
X-Ray photoelectron spectroscopy (XPS) was performed in an AXIS Supra XPS by Kratos Analytical. XPS spectra were collected using a monochromatized Al Kα radiation (hυ = 1486.7 eV) under a base pressure of $10^{-9}$ Torr. To avoid moisture and air exposure, a nitrogen filled glovebox was directly connected to XPS spectrometer. All XPS measurements were collected with a 300 × 700 μm² spot size. Survey scans were performed with a step size of 1.0 eV, followed by a high-resolution scan with 0.1 eV resolution, for lithium 1s, carbon 1s, oxygen 1s, nitrogen 1s, and phosphorous 2p regions. A 5 keV Ar plasma etching source was used for depth profiling with a pre-etching for 5 s, etching for 60 s and post-etching for 10 s. All spectra were calibrated with adventitious carbon 1s (284.6 eV) and analyzed by CasaXPS software.

**Cryogenic X-ray Diffraction**
The powder crystal X-ray diffraction was carried out on a Bruker micro focused rotating anode, with double bounced focusing optics resulting in Cu $K_{\alpha 1}$ and $K_{\alpha 2}$ radiation ($\lambda_{avg}$ =1.54178 Å ) focused at the sample. A sample of LiPON was mounted onto a four circle Kappa geometry goniometer with APEX II CCD detector. The sample was cooled and data were collected in a nitrogen gas stream at 100 K.

**Electron Energy Loss Spectroscopy Simulation**
Electron energy loss spectroscopy simulations were conducted using FEFF9 software. The crystal structures used included a $Li_2O$ cif file (ID #22402), a $Li_3PO_4$ cif file (ID # 77095) and a $Li_3P$ cif file (ID # 240861) taken from ICSD database. The amorphous LiPON structure was generated by AIMD. The simulation parameters for FEFF9 included beam energy of 200 keV, collection and convergence angles of 10 mrads, xkmax value of 4, xkstep value of 0.02 and estep value of 0.01. Hedin Lundqvist exchange and RPA corehole were used for electron core interactions in FEFF9.

**Calculation of Diffusion Barriers in Li Metal**
Density functional theory (DFT) calculations were performed using the generalized gradient approximation (GGA)[52] approximation, and projector augmented-wave method (PAW)[53] pseudopotentials were used as implemented by the Vienna Ab initio Simulation Package (VASP)[54,55]. The Perdew-Burke-Ernzerhof exchange correlation[56] and a plane wave representation for the wavefunction with a cutoff energy of 450 eV were used. For calculations of diffusion in Li metal, the Brillouin zone was sampled with a k-point mesh of 5x5x5 for both structural relaxations and nudged elastic band (NEB)[57] calculations. NEB calculations were performed placing dopant ions in interstitial locations of a 128 Li atom unit cell and interpolating 5 intermediate images.

## Acknowledgements

The authors gratefully acknowledge funding support from the U.S. Department of Energy, Office of Basic Energy Sciences, under Award Number DE-SC0002357 (program manager Dr. Jane Zhu). FIB was performed at the San Diego Nanotechnology Infrastructure (SDNI), a member of the National Nanotechnology Coordinated Infrastructure, which is supported by the National Science Foundation (Grant ECCS1542148). TEM and XPS were performed at the UC Irvine Materials Research Institute (IMRI). XPS work was performed at the UC Irvine Materials Research Institute (IMRI) using instrumentation funded in part by the National Science Foundation Major Research Instrumentation Program under grant no. CHE-1338173. This work also used the Extreme Science and Engineering Discovery Environment (XSEDE), which is supported by National Science Foundation grant number ACI-1548562.

## Author contributions

D.C., T.A.W., X.W. and Y.S.M conceived the ideas. D.C. and T.A.W. developed the cryo-lift-out methodology. D.C. prepared thin film sample, cryo FIB sample and TEM sample with the help of T.A.W. and R.S. R.S. fabricated the thin film full cell and conducted electrochemical cycling. D.C. and S.W. designed and performed XPS experiments. X.W., D.C., C. F and B.L. designed and conducted cryo-TEM and EDS line scan. M.Z., S.B. and D.C. conducted cryo-STEM and EELS measurements. D.C., X.W., T.A.W. and M.Z. interpreted (S)TEM data. T.A.W. performed DFT calculations. D.C. performed FEFF9 simulation. D.C., T.A.W. and M.Z. interpreted the FEFF9 simulation results. H.N. and D.C. collected cryo-XRD data. D.C and M. K. generated the interphase schematic and cryo-lift-out schematic. W.L. fabricated the liquid electrolyte cell and conducted electrochemical cycling. D.C., T.A.W., X.W., S.J.K and Y.S.M. co-wrote the manuscript. All authors discussed the results and commented on the manuscript. All authors have approved the final manuscript.

## Declaration of Interests

The authors declare no competing interests.

**Supplemental Information**

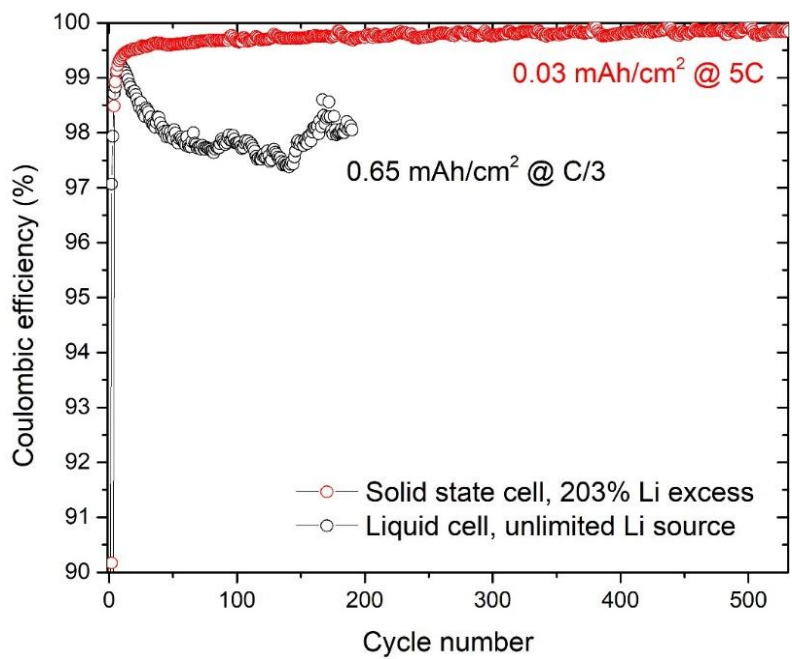

**Supplementary Figure 1. Electrochemical performance of thin film solid state cell and liquid cell using Gen2 electrolyte (EC:EMC=3:7 wt%, 1M LiPF$_6$).** Note that the Columbic efficiency of liquid cell is hard to stabilize beyond 99% due to the continuous electrolyte decomposition during charging process, while thin film solid state cell can cycle stably at a Columbic efficiency beyond 99.85% since LiPON can withstand the high oxidative potential.

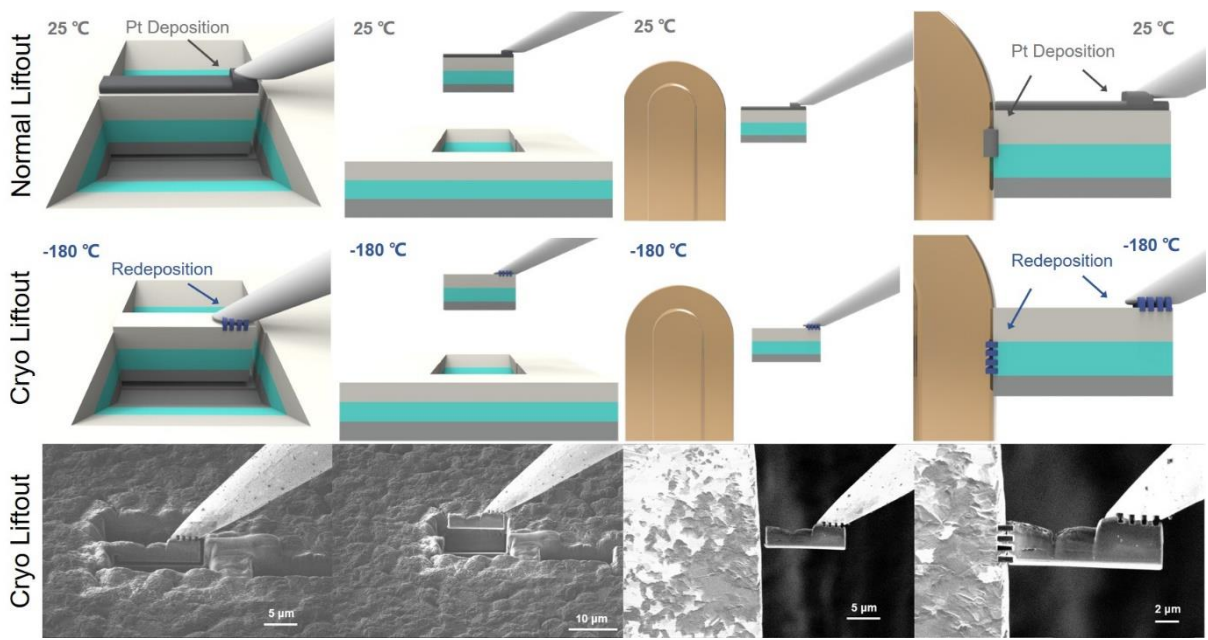

**Supplementary Figure 2. Redeposition mounting methodology** During TEM sample preparation in FIB. after cross-section milling, cleaning and J-cutting, the lamella needs to be connected with the tungsten probe for liftout process. Normal method is depositing Pt as the connection at room temperature (25℃), which, however, cannot be realized at liquid nitrogen temperature. Regarded as a side effect of ion milling in most cases, redeposition of sputtered material is common during FIB milling processes. Here we introduce a new methodology for lamella lift-out under cryogenic conditions, where the key is the redeposition of sputtered material. As shown in Figure S2, at -180℃, several rectangular milling patterns are drawn at the junction of tungsten probe and lamella top surface. A 10-pA ion beam current is then used to mill through the patterned region, where the redeposition materials will redeposit at the surrounding region and connect lamella with the tungsten probe. The lamella can then be lifted out by the tungsten probe without any Pt deposition after cutting free. As the lamella is in contact with the Cu grid post, same method is applied again where several rectangular milling patterns are milled through to let reposition connect the lamella with the Cu grid post. By using this destructive method, we constructively complete the lamella lift-out fully under cryogenic conditions, successfully maintain the morphology of Li metal, and preserve the Li/LiPON interphase for further characterizations.

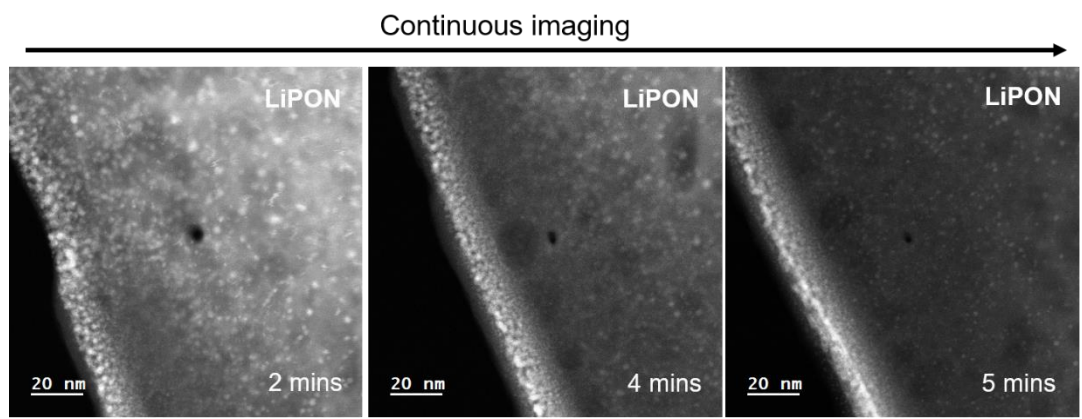

**Supplementary Figure 3. Beam stability demonstration of LiPON under high-magnification cryo-STEM**

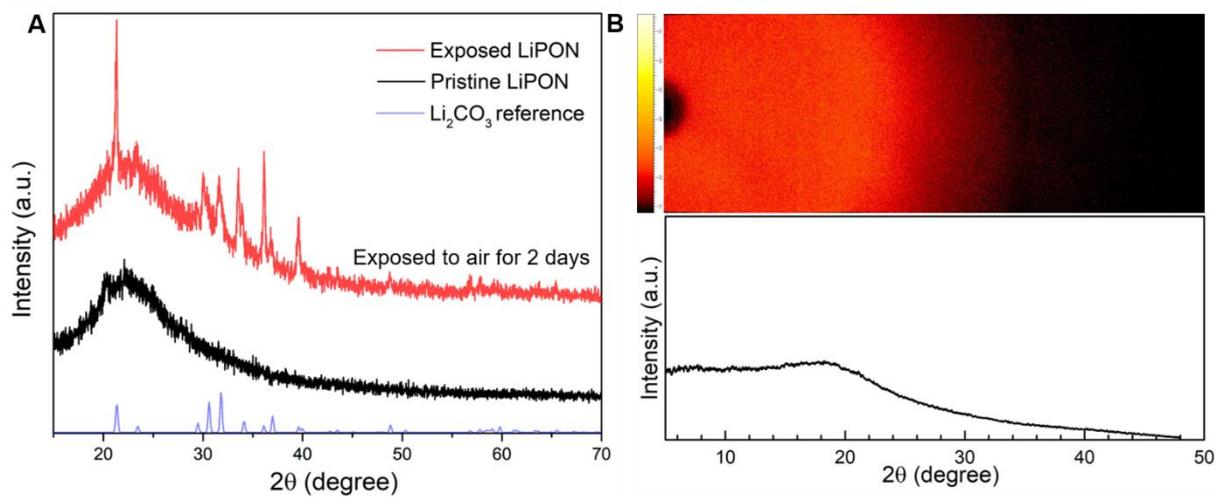

**Supplementary Figure 4. Room-temperature and cryo-temperature XRD of LiPON** (A) The XRD pattern of pristine LiPON (black) and LiPON exposed in air for 2 days (red). The peaks on the exposed LiPON can be indexed to the peaks from $Li_2CO_3$ as shown at the bottom. (B) The cryo-XRD pattern of pristine LiPON showing the amorphous phase of LiPON at 100 K.

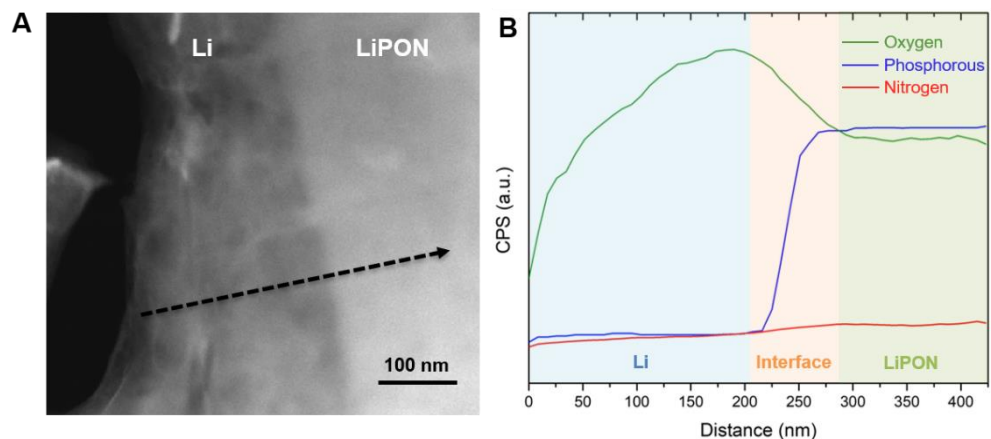

**Supplementary Figure 5. EDS linescan at the interphase.** (A) Cryo STEM DF image of Li/LiPON interphase. (B) EDS linescan CPS of O, P and N signals with respect to distance along the black dashed arrow in (A). The Li metal region, interphase region and LiPON region are indicated by the blue, orange and green background, respectively. The interphase region is defined with a length of 76 nm based on the CPS change of P and N. Note that it is common to observe extraneous O signal for the elemental analysis in STEM/EDS, which may come from the grid and holder. It is unlikely to have O impurities during Li deposition since the Li deposition was performed in a high-vacuum chamber installed in an ultra-clean glovebox with <0.1 ppm $O_2$ and none of any organic solvent.

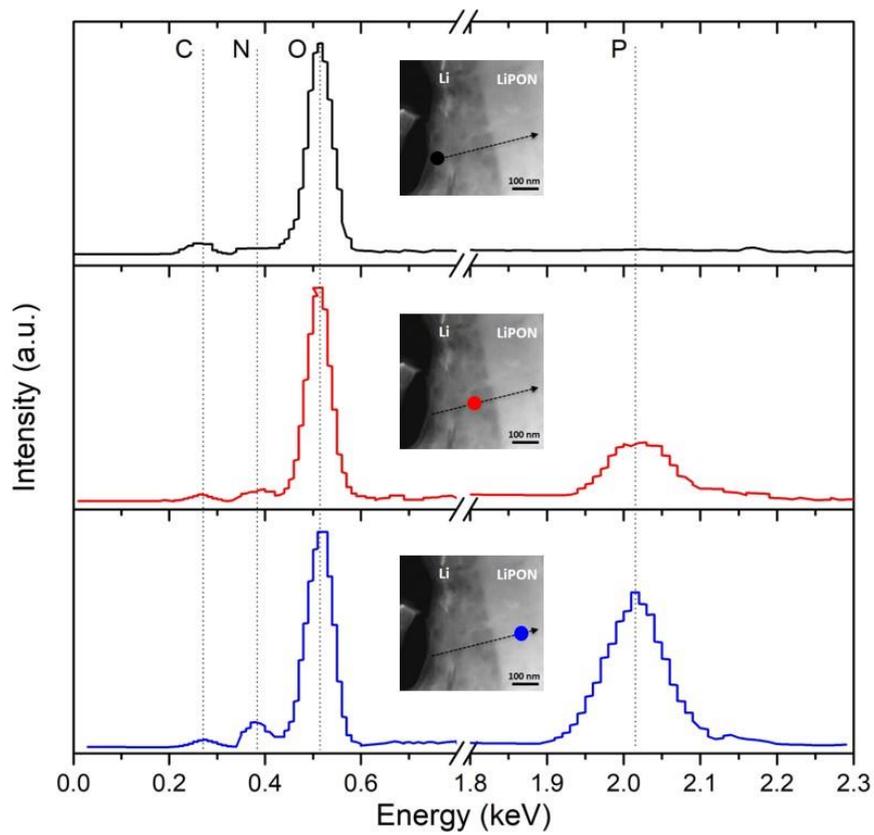

**Supplementary Figure 6. EDS spectra evolution along the interface.** Three spots were selected to elaborate the EDS spectra evolution from the beginning of the interphase (black), at the interphase (red) and till the end of the interphase (blue).

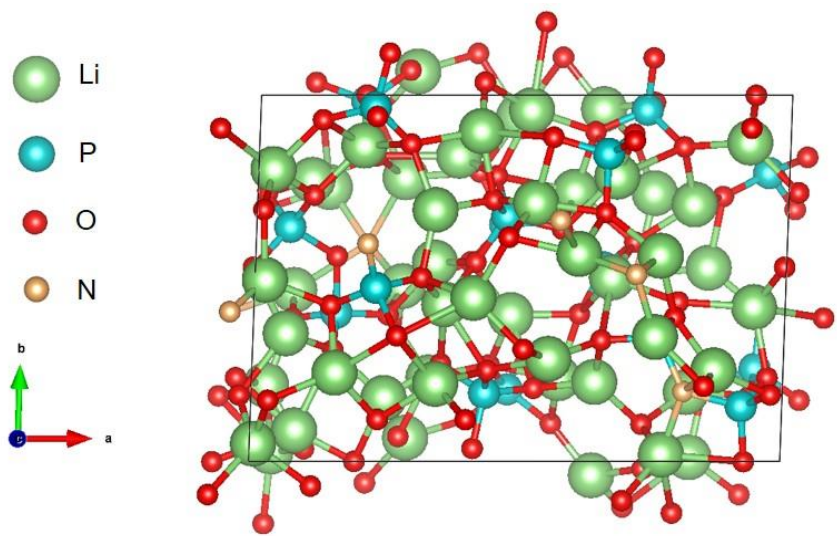

**Supplementary Figure 7. Amorphous LiPON structure generated by AIMD.** This structure contains 46 Li atoms, 16 P atoms, 55 O atoms and 5 N atoms, with a stoichiometry of $Li_{2.88}PO_{3.44}N_{0.31}$.

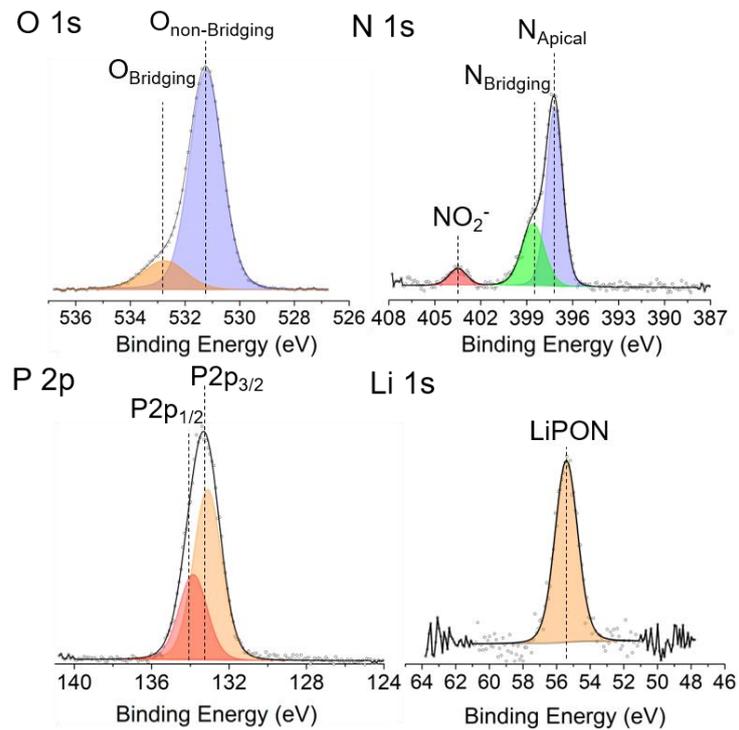

**Supplementary Figure 8. Reference XPS spectra of LiPON for O 1s, N 1s, P 2p and Li 1s regions** O 1s region is fitted to non-bridging O and bridging O peaks located at 531.3 eV and 532.7 eV, respectively. N 1s region is fitted to apical N, bridging N and $NO_2^-$ species located at 397.3 eV, 398.5 eV and 403.7 eV, respectively. P 2p region is fitted to P $2p^{3/2}$ and P $2p^{1/2}$ peaks at 133.2 eV and 133.8 eV, respectively. Li 1s region is assigned with the signal of Li from LiPON. The as-deposited LiPON has an approximate stoichiometry of $Li_{2.84}PO_{3.35}N_{0.38}$.

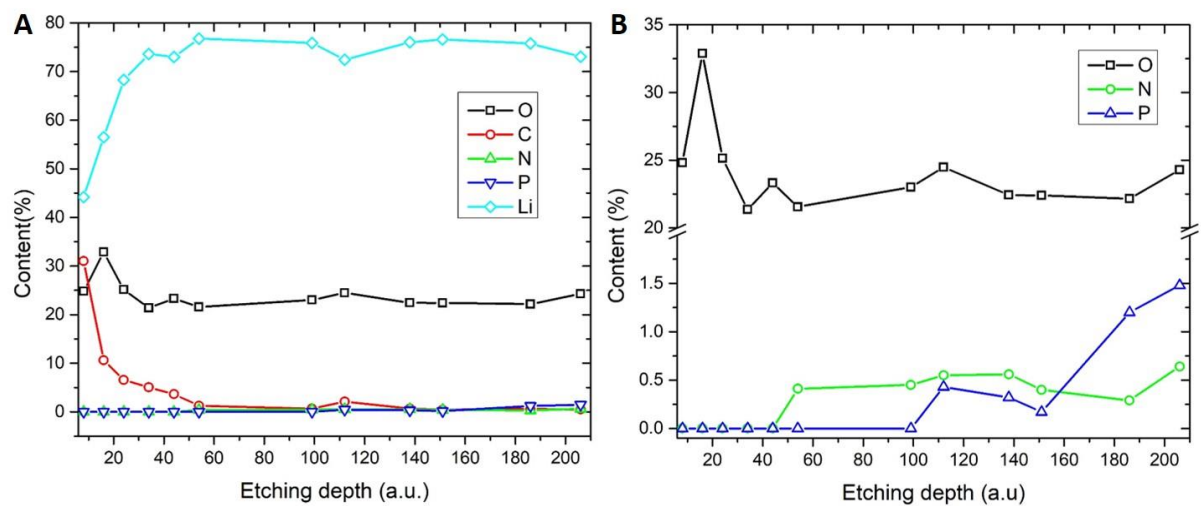

**Supplementary Figure 9. Chemical evolution of O 1s, N 1s, P 2p and Li 1s along Li/LiPON interphase by XPS depth profiling.** (A) Composition content evolution of O, C, N, P and Li elements through the interphase. (B) zoomed-in content evolution plot of O, N and P elements from (A), where gradients of N and P are observed and N signal appears first at the interphase.

| Transfer Process | Protection Method |
|---|---|
| Bulk sample from glovebox to FIB chamber | Inert gas environment (Ar) |
| TEM sample from FIB chamber to glovebox | High vacuum |
| TEM sample from glovebox to TEM holder | Inert gas environment (Ar)/ cryogenic temperature |
| TEM sample from TEM holder to TEM column | Inert gas environment (Ar)/ cryogenic temperature |

**Supplementary Table 1. Protection methods for each sample transfer process.**

| | Electronic Conductivity (S/cm) at 298K | Ionic Conductivity (S/cm) at 298K |
|---|---|---|
| **$Li_2O$**[1] | $10^{-14}$ | $10^{-12}$ |
| **$Li_3N$**[2] | $10^{-12}$ | $10^{-4}$ |
| **$Li_3PO_4$**[3,4] | $10^{-9}$ | $10^{-8}$ |
| **$LiF$**[5,6] | $<10^{-12}$ | $<10^{-8}$ |
| **$LiCl$**[7,8] | $10^{-6}$ | $<10^{-7}$ |
| **$Li_2S$**[9,10] | $10^{-9}$ | $10^{-8}$ |

**Supplementary Table 2. Electrical and ionic conductivities of different SEI components**